\documentclass[10pt]{iopart} 
\begin{document}
\title[Necessary and sufficient conditions for existence of bound states]{Necessary and sufficient conditions for existence of bound states in a central potential}
\author{Fabian Brau\footnote[1]{E-Mail: fabian.brau@umh.ac.be}}
\address{Service de Physique G\'en\'erale et de Physique des Particules El\'ementaires, Groupe de Physique Nucl\'eaire Th\'eorique, Universit\'e de Mons-Hainaut, Mons, Belgique}
\date{\today}

\begin{abstract}
We obtain, using the Birman-Schwinger method, a series of necessary conditions for the existence of at least one bound state applicable to arbitrary central potentials in the context of nonrelativistic quantum mechanics. These conditions yield a monotonic series of lower limits on the ``critical" value of the strength of the potential (for which a first bound state appears) which converges to the exact critical strength. We also obtain a sufficient condition for the existence of bound states in a central monotonic potential which yield an upper limit on the critical strength of the potential.
\end{abstract}

\maketitle

\section{Introduction}
\label{sec1}

The problem of finding upper and lower limits on the number of bound states of a given potential  is become a classical problem since the pioneer works of Jost and Pais in 1951 \cite{jost51} and Bargmann in 1952 \cite{barg52}. They obtained, for the first time, a necessary condition for the existence of bound states in a central potential which can be obtained from the following upper limit on the number of $\ell$-wave bound states (setting $N_{\ell}$ to 1)
\begin{equation}
\label{eq1}
N_{\ell}\le \frac{1}{2\ell+1}\int_0^{\infty}dr\, r\, |V^-(r)|.
\end{equation}
In this inequality, $V^-(r)$ is the negative part of the potential obtained by setting its positive part to zero and $\ell$ is the angular momentum. Note that we use the standard quantum-mechanical units such as $\hbar=2m=1$, where $m$ is the mass of the particle. This upper limit (\ref{eq1}), called Bargmann-Schwinger upper limit in the literature, was the starting point of intensive studies and a fairly large number of upper and lower limits on the number of bound states for various class of potentials was found, see for example \cite{bi61,sc61,ca65a,ca65b,ca65c,ch68,ma72,gl76,si76,ma77,li80,ch95a,ch95b,ch96,bl96,la97,br03a,br03b,br03c}.

An important theorem for classifying these results was found by Chadan \cite{ch68} and gives the asymptotic behavior of the number of bound states as the strength, $g$, of the central potential goes to infinity:
\begin{equation}
\label{eq2}
N\approx \frac{g^{1/2}}{\pi}\int_0^{\infty}dr\, v(r)^{1/2}\quad {\rm as}\quad g\rightarrow \infty,
\end{equation}
where the symbol $\approx$ means asymptotic equality and $V^-(r)=-g\, v(r)$. This result implies that any upper and lower limit from which could yield cogent results should behave asymptotically as $g^{1/2}$. More importantly, the relation (\ref{eq2}) gives the functional of the potential, that is to say the coefficient in front of $g^{1/2}$, that appears in the asymptotic behavior. The upper limit (\ref{eq1}) is proportional to $g$ instead of $g^{1/2}$ and is not very stringent for strong potentials. Upper and lower limits featuring the correct $g^{1/2}$ dependency was first obtained in the Ref. \cite{ca65c}. Upper and lower limits featuring the correct asymptotic behavior (\ref{eq2}) was first derived in Refs. \cite{br03a,br03b}. In practice, the asymptotic regime is reached very quickly when the strength of the potential is large enough to bind two or three bound states.

The situation is completely different when one consider the transition between 0 and 1 bound state and in particular upper and lower limit on the ``critical" value of the strength of the potential, $g_{{\rm c}}$, for which a first bound state appears. In this case, there is no theorem to know in advance which limit yield the most stringent restriction on $g_{{\rm c}}$. It is then of interest to obtain various limits, since the limit yielding the most stringent restriction change from one potential to another.

In Section \ref{sec2}, we obtain a series of necessary conditions for the existence of at least one bound state, applicable to arbitrary central potentials, which converges to the exact critical strength. In Section \ref{sec3}, we present a sufficient condition for the existence of bound states in a central monotonic potential. In Section \ref{sec4}, we perform several tests of the cogency of the limits presented in this article and we compare them to some previously known results and to the exact results.

\section{Necessary conditions}
\label{sec2}

The necessary conditions for the existence of bound states derived in this section is obtained with the help of a simple extension of the Birman-Schwinger method. Birman \cite{bi61} and Schwinger \cite{sc61} have shown how to obtain an upper limit on the number of bound states once the Green function of the kinetic energy operator of a wave equation is known. We recall briefly the main line of the method applied to the radial Schr\"odinger equation for completeness; for more details see the original articles \cite{bi61,sc61}.

The Schr\"odinger equation for a central potential $V(r)$ reads
\begin{equation}
\label{eq3}
\left(-\frac{d^2}{dr^2}+\frac{\ell(\ell+1)}{r^2}\right)u_{\ell}(r)=(E-V(r))\, u_{\ell}(r).
\end{equation}
The zero-energy Schr\"odinger equation can be written under the form of the following integral equation
\begin{equation}
\label{eq4}
u_{\ell}(r)=-\int_0^{\infty}dr'\, g_{\ell}(r,r')\, V(r')\, u_{\ell}(r'),
\end{equation}
where $g_{\ell}(r,r')$ is the Green function of the kinetic energy operator and is explicitely given by
\begin{equation}
\label{eq5}
g_{\ell}(r,r')=\frac{1}{2\ell+1} r_<^{\ell+1}\, r_>^{-\ell},
\end{equation}
where $r_<=\min[r,r']$ and $r_>=\max[r,r']$.
Since the purpose of the method is to obtain an upper limit on the number of bound states, we can replace $V(r)$ by $-|V^-(r)|$ where $V^-(r)$ is the negative part of the potential obtained by setting the positive part of the potential equal to zero. Indeed, a decrease of the potential in some region must lower the energies of the bound states and therefore cannot lessen their number. Moreover, we introduce the parameter $0<\lambda\le1$ by the substitution $|V^-(r)|\rightarrow \lambda|V^-(r)|$. As $\lambda$ increases from 0, we reach a critical value, $\lambda_1$, at which a bound state first appears with a vanishing binding energy, $E=0$. With further growth of $\lambda$, the energy of this state decreases until we reach a second critical  value, $\lambda_2$, at which a second bound state appears and so on. When $\lambda$ has attained the value unity and, $\lambda_{N_{\ell}}\le 1<\lambda_{N_{\ell}+1}$, there are $N_{\ell}$ bound states.

We now introduce, to obtain a symmetrical kernel, a new wave function as
\begin{equation}
\label{eq6}
\phi_{\ell}(r)=|V^-(r)|^{1/2}\, u_{\ell}(r).
\end{equation}
The equation (\ref{eq4}) becomes
\begin{equation}
\label{eq7}
\lambda^{-1}\,\phi_{\ell}(r)=\int_0^{\infty}dr'\, K_{\ell}(r,r')\, \phi_{\ell}(r'),
\end{equation}
where $K_{\ell}(r,r')$ is given by
\begin{equation}
\label{eq8}
K_{\ell}(r,r')=|V^-(r)|^{1/2}\, g_{\ell}(r,r')\,|V^-(r')|^{1/2}.
\end{equation}
The kernel being positive, we have $0<\lambda_1<\lambda_2<\cdots<\lambda_N\le 1$ and
$0<\lambda_k<\infty$ ($\lambda_k$ denotes each eigenvalue of (\ref{eq7})). It is well known that the trace of the iterated kernels equals the sum of the eigenvalues of the integral equation (\ref{eq7}) as follow
\begin{equation}
\label{eq9}
\sum_{k=1}^{\infty} \frac{1}{(\lambda_k)^n} = \int_0^{\infty} dr\, K_{\ell}^{(n)}(r,r),
\end{equation}
where the iterated kernel $K_{\ell}^{(n)}(s,t)$ is given by
\begin{equation}
\label{eq10}
K_{\ell}^{(n)}(s,t) = \int_0^{\infty} du\ K_{\ell}(s,u)\, K_{\ell}^{(n-1)}(u,t),
\end{equation}
with 
\begin{equation}
\label{eq11}
K_{\ell}^{(1)}(s,t) \equiv K_{\ell}(s,t),
\end{equation}
and $n=1,2,\ldots$. Now it is plain that the following inequalities hold
\begin{equation}
\label{eq12}
\sum_{k=1}^{\infty} \frac{1}{(\lambda_k)^n}\ge \sum_{k=1}^{N_{\ell}} \frac{1}{(\lambda_k)^n}>N_{\ell},
\end{equation}
where $N_{\ell}$ is the number of $\ell$-wave bound states. From (\ref{eq9}), (\ref{eq10}), (\ref{eq11}) and (\ref{eq12}) we find that an upper limit on the number of $\ell$-wave bound states of the Schr\"odinger equation is given by
\begin{equation}
\label{eq13}
N_{\ell}<\int_0^{\infty} dr\, K_{\ell}^{(n)}(r,r).
\end{equation}

In his article, Schwinger consider only the case $n=1$ for the equation (\ref{eq13}) which yields the Bargmann-Schwinger upper limit (\ref{eq1}). Indeed, greater values of $n$ would yield upper limits which possess a worse dependency on the strength of the potential $g$ than the upper limit (\ref{eq1}) and which would be very poor for strong potentials. But it appears that, as illustrated in Section \ref{sec4}, the larger is the value $n$ the better is the lower limit on the critical value of strength of the potential. 

The necessary conditions for the existence of $\ell$-wave bound states obtained from (\ref{eq13}) read respectively for $n=1,2,3$:
\begin{equation}
\label{eq14}
\frac{1}{2\ell+1}\int_0^{\infty} dr\, r\, |V^-(r)|\ge 1, 
\end{equation}
\begin{equation}
\label{eq15}
\frac{2}{(2\ell+1)^2}\int_0^{\infty} dr_1\, r_1^{-2\ell}\, |V^-(r_1)|\int_0^{r_1} dr_2\, r_2^{2\ell+2}\, |V^-(r_2)|\ge 1,
\end{equation}
\begin{eqnarray}
\label{eq16}
&&\frac{6}{(2\ell+1)^3}\int_0^{\infty} dr_1\, r_1^{-2\ell}\, |V^-(r_1)|\int_0^{r_1}dr_2\, r_2\, |V^-(r_2)| \nonumber \\
&&\times \int_0^{r_2} dr_3\, r_3^{2\ell+2}\, |V^-(r_3)|\ge 1,
\end{eqnarray}
The improvements of the lower limits on $g_{{\rm c}}$ implied by the relations (\ref{eq15}) and (\ref{eq16}) over the lower limit inferred from the well known relation (\ref{eq14}) are illustrated in Section \ref{sec4} for a  square-well potential and an exponential potential.

Let us end this section by noting that the procedure employed here yield also a necessary condition for the existence of bound states analogous to the condition obtained by Glaser {\it et al.} \cite{gl76}
\begin{equation}
\label{eq24}
\frac{(p-1)^{p-1}\Gamma(2p)}{(2\ell+1)^{2p-1}\, p^p\Gamma^2(p)}\, \int_0^{\infty}dr\, r^{2p-1}\, 
|V^-(r)|^p\ge 1,
\end{equation}
where $p>1$ must be chosen to optimize the result. Indeed, for $\ell>0$, we can use the $n$ times the H\"older inequality in the relation (\ref{eq13}) and taking $n$ going to infinity (see \cite{br03d} for more details) we obtain
\begin{equation}
\label{eq17}
\left[\frac{(2\ell+1)\, p(p-1)}{p^2(\ell+2)(\ell-1)+3p-1}\right]^{p-1}\int_0^{\infty} dr\, r^{2p-1}\, |V^-(r)|^p\ge 1.
\end{equation}
The constant in front of the integral is unfortunately always greater than the constant appearing in the necessary condition (\ref{eq24}), and the relation (\ref{eq17}) is thus always less stringent.

\section{Sufficient condition}
\label{sec3}

The sufficient condition is obtained with the help of a generalization of the comparison theorem proved recently and where the comparison potentials intersect (Theorem 7 of Ref. \cite{hall02}). The new theorem reads

{\bf Theorem.} {\it If two monotonic potentials $V_1(r)$ and $V_2(r)$ cross twice for $r>0$ at $r=r_1,r_2$ $(r_1<r_2)$ with}
\begin{eqnarray}
&&(i)\quad  V_1(r)<V_2(r)\quad for \quad 0<r<r_1 \quad and \nonumber \\
&&(ii)\quad \int_0^{r_2}dy\,[V_1(y)-V_2(y)]\, y^2\le 0 \nonumber, 
\end{eqnarray}
{\it then $E_1<E_2$, where $E_{1,2}$ are the ground states of the potentials $V_{1,2}(r)$.}

As comparison potential $V_2(r)$, we choose a simple square-well
\begin{equation}
\label{eq18}
V_2(r)=-V_0 \, \theta(R-r),
\end{equation}
where $\theta(x)$ is the Heaviside function. Moreover, we choose this potential such as a zero-energy bound state exists: $V_0 R^2=\pi^2/4$. This implies that the potential $V_1(r)$ possesses at least one bound state. For this particular choice of $V_2(r)$ we have $r_2=R$. We write the potential $V_1(r)$ under the form 
\begin{equation}
\label{eq18b}
V_1(r)=-g s^{-2}\, v(r/s,k),
\end{equation}
where $k$ are the other parameters of the potential. The hypothesis ({\it ii}) above yields the following upper bound $g^{\rm{up}}_{\rm{c}}$ on the critical coupling constant $g_{\rm{c}}$ 
\begin{equation}
\label{eq19}
g^{\rm{up}}_{\rm{c}}=\frac{\pi^2}{12}\, \frac{\alpha}{\int_0^{\alpha} dy\, y^2\, v(y,k)},
\end{equation}
where $\alpha =R/s$. The best restriction is obviously obtained with the value of $\alpha$  minimizing the right-hand side of (\ref{eq19}). The upper limit can thus be written as
\begin{equation}
\label{eq20}
g^{\rm{up}}_{\rm{c}}=\frac{\pi^2}{12}\, \frac{1}{\alpha^2\, v(\alpha,k)},
\end{equation}
where $\alpha$ is the unique solution of
\begin{equation}
\label{eq21}
\int_0^{\alpha} dy\, y^2\, v(y,k)=\alpha^3 v(\alpha,k).
\end{equation}
The definition (\ref{eq21}) of $\alpha$ has a simple geometric signification which allows to remark that $\alpha>\max[y^2 v(y,k)]$.

Obviously, we have used a very particular comparison potential $V_2(r)$ to write a neat formula for the upper limit on the critical coupling constant $g_{\rm{c}}$. In practice, a better upper limit could be obtained by the use of a more appropriate comparison potential for which the exact value of the critical coupling constant is known (and for which the conditions ($i$) and ($ii$) apply!).

\section{Tests}
\label{sec4}

The first potential we consider to test the limits presented in the previous sections is a square-well potential that we write in the convenient form
\begin{equation}
\label{eq22}
V(r)=-gR^{-2}\, \theta(1-r/R).
\end{equation}
The sufficient condition (\ref{eq20})-(\ref{eq21}), applicable only for $\ell=0$, is saturated for this potential (with $\alpha=1$) and thus leads to the exact result. The necessary conditions (\ref{eq14})-(\ref{eq16}) give the following lower limits
\begin{eqnarray}
\label{eq23a}
g_{\rm{c}}^{\rm{lo}}=2(2\ell+1), \\ \label{eq23b}
g_{\rm{c}}^{\rm{lo}}=(2\ell+1)[2(2\ell+3)]^{1/2}, \\ \label{eq23c}
g_{\rm{c}}^{\rm{lo}}=(2\ell+1)[(2\ell+3)(2\ell+5)]^{1/3}.
\end{eqnarray}
The comparison between the new lower limits on $g_{{\rm c}}$, the limit (\ref{eq24}) and the exact results is reported in Table \ref{tab1} and shows that the new limits are quite cogent and converge quickly to the exact result especially for small value of $\ell$.

\begin{table}
\protect\caption{Comparison between the exact values of the critical coupling constant 
$g_{{\rm c}}$ of a square well potential for various value of $\ell$ and the lower limits, $g_{{\rm c}}^{{\rm lo}}\le g_{{\rm c}}$, obtained with the relations (\ref{eq23a})-(\ref{eq23c}), the lower limit obtained with the relation (\ref{eq13}) with $n=4$ and $N_{\ell}=1$ (calculated numerically) and the lower limit obtained with the formula (\ref{eq24}) (with the optimal value of $p$).}
\label{tab1}
\begin{center}
\begin{tabular}{ccccccc}
\hline
$\ell$ & $n=1$ & $n=2$ & $n=3$ & $n=4$ & Eq. (\protect\ref{eq24}) & Exact \\
\hline
0 &  2   &  2.4495  &  2.4662   &  2.4672  &  2.3593  &  2.4674     \\
1 &  6   &  9.4868  &  9.8132   &  9.8592  &  9.1220  &  9.8696     \\
2 &  10  &  18.708  &  19.895   &  20.120  &  18.454  &  20.191     \\
3 &  14  &  29.699  &  32.383   &  32.981  &  30.245  &  33.217     \\
4 &  18  &  42.214  &  47.064   &  48.272  &  44.425  &  48.831     \\
5 &  22  &  56.089  &  63.788   &  65.868  &  60.947  &  66.954     \\
\hline
\end{tabular}
\end{center}
\end{table}

The last test is performed with an exponential potential written as
\begin{equation}
\label{eq25}
V(r)=-gR^{-2}\, \exp(-r/R).
\end{equation}
For $\ell=0$, the sufficient condition (\ref{eq20})-(\ref{eq21}) leads to $g_{\rm{c}}^{\rm{up}}=2.118$ while the exact result is given by $g_{\rm{c}}=z_0^2/4\cong 1.4458$ ($z_0=2.4048$ is the first zero of the Bessel function $J_0(x)$). The upper limit is not very stringent for this potential because the comparison potential that we choose (a square well) is very different from an exponential potential. The upper limit yields more cogent result, for example, for a Wood-Saxon potential. For an exponential potential a better upper limit can be obtained with the Calogero lower bound \cite{ca65a}: $g_{{\rm c}}^{{\rm up}}=1.677$.

\begin{table}
\protect\caption{Comparison between the exact values of the critical coupling constant 
$g_{{\rm c}}$ of an exponential potential for various value of $\ell$ and the lower limits, $g_{{\rm c}}^{{\rm lo}}\le g_{{\rm c}}$, obtained with the relations (\ref{eq23a})-(\ref{eq23c}), the lower limit obtained with the relation (\ref{eq13}) with $n=4$ and $N_{\ell}=1$ (calculated numerically) and the lower limit obtained with the formula (\ref{eq24}) (with the optimal value of $p$).}
\label{tab2}
\begin{center}
\begin{tabular}{ccccccc}
\hline
$\ell$ & $n=1$ & $n=2$ & $n=3$ & $n=4$ & Eq. (\protect\ref{eq24}) & Exact \\
\hline
0 &  1   &  1.4142  &  1.4422   &  1.4453  &  1.4383  &  1.4458     \\
1 &  3   &  6.2700  &  6.8546   &  6.9913  &  7.0232  &  7.0491     \\
2 &  5   &  13.145  &  15.257   &  15.804  &  16.277  &  16.313     \\
3 &  7   &  21.593  &  26.265   &  27.364  &  29.218  &  29.259     \\
4 &  9   &  31.363  &  39.616   &  41.296  &  45.849  &  45.893     \\
5 &  11  &  42.297  &  55.120   &  57.480  &  66.173  &  66.219     \\
\hline
\end{tabular}
\end{center}
\end{table}

The comparison between the news lower limits on $g_{{\rm c}}$, the limit (\ref{eq24}) and the exact result is reported in the Table \ref{tab2}. The new lower limits on $g_{\rm{c}}$ are quite cogent and converge quickly to the exact results especially for small value of $\ell$, but this convergence is slower than in the case of a square well potential.

\ack

We would like to thank the FNRS for financial support (FNRS Postdoctoral Researcher position).

\end{document}